\documentstyle[mprocl]{article}

\def\cA{{\cal A}}

\def\be{\begin{equation}}
\def\ee{\end{equation}}
\def\bea{\begin{eqnarray}}
\def\eea{\end{eqnarray}}

\begin{document}

\title{Einstein Gravity --- Supergravity Correspondence}

\author{Chiang-Mei Chen}

\address{Department of Physics, National Central University, Chungli \\
         E-mail: cmchen@joule.phy.ncu.edu.tw}

\author{Dmitri V. Gal'tsov and Sergei A. Sharakin}

\address{Department of Theoretical Physics, Moscow State University,
         Moscow \\ E-mail: galtsov@grg.phys.msu.su,
         sharakin@grg1.phys.msu.su}

\maketitle\abstracts{
A correspondence between the three-block truncated $11D$ supergravity
and the $8D$ pure Einstein gravity with two commuting Killing
symmetries is discussed.
The Kaluza-Klein two-forms of the $6D$ theory obtained after dimensional
reduction along the Killing orbits generate the four-form
field of supergravity via an inverse dualization.
Thus any solution to the vacuum Einstein equations in eight dimensions
depending on six coordinates have $11D$-supergravity counterparts
with the non-trivial four-form field.
Using this proposed duality we derive a new dyon solution
of 11D supergravity  describing the $M2$ and $M5$--branes
intersecting at a point.}

\section{Introduction}
Recently the embedding of the classical solutions of lower dimensional
(truncated) gauged supergravities into the type IIB string and M--theory
has attracted much attention as providing a simple way to find
solutions of the latter theories in terms of lower dimensional
models. This type of embedding is based on the $S^n$
(transversal space of desired solution) compactifications as well as a
consistent truncation of scalar fields and vector multiplets.
For examples, the charged $AdS_5$, $AdS_4$ and $AdS_7$ black holes
had been embedded as the $D3$--branes, $M2$--branes and $M5$--branes
respectively. It is worth noting that the charges of the $AdS$
black holes may be interpreted as the
angular momenta of the associated branes \cite{CvDuHo99}.

Here we suggest another way of embedding \cite{ChGaSh99a} of
lower-dimensional gravity solutions into eleven-dimensional
supergravity using a non-local duality between solutions in the vacuum
Einstein theory in eight dimensions and $11D$ supergravity.
This symmetry is based on the
fact that the six-dimensional reduction of the eight-dimensional vacuum
Einstein theory in presence of two spacelike commuting Killing vectors
admits, after suitable field redefinitions, the same form as
the truncated boson sector of $D=11$ supergravity compactified on $T^5$
with some additional restrictions on the metric and the three--form field.
This correspondence provides a new scheme of uplifting
the $D \le 8$ dimensional solutions to vacuum Einstein gravity,
especially the non-linear superpositions
of known Kaluza--Klein solutions such as pp-waves, KK-monopole,
Melvin universes etc.,
to their supergravity counterparts, producing, for instance,
intersecting $M$--fluxbranes \cite{ChGaSh99c} as well as a new type
composite $M2 \cup M5$--brane \cite{ChGaSh99a} corresponding to the
Kaluza--Klein dyon black hole.

The proposed new mapping from $8D$ to $11D$ is `one to many',
depending on the choice of two Killing vectors in eight dimensions.
This opens a way to proliferate $11D$ solutions taking as a seed
an already known one, lowering it down to eight dimensions according to
the parameterization suggested, and then coming back to eleven dimensions
with a different choice of (ordered) Killing vectors. This generates the
whole class of supergravity solutions related non-locally to each other.

\section{11D Supergravity vs. 8D Einstein Gravity}
The bosonic sector of $11D$ supergravity contains the eleven-dimensional
metric $\hat g_{AB}$ and a three-form gauge field $\hat A_{[3]}$ described
by a Lagrangian which possesses a non-vanishing Chern-Simons term \cite{St97}
\be
S_{11} = \int d^{11}x \sqrt{-\hat g_{11}} \left\{ \hat R_{11}
   - \frac1{48} \hat F^2_{[4]} \right\} \nonumber\\
   - \frac16 \int \hat F_{[4]} \wedge \hat F_{[4]} \wedge\hat A_{[3]},
      \label{11d}
\ee
where $\hat F_{[4]} = d \hat A_{[3]}$.
Assuming the space--times can be decomposed into three--block as
$M^{11} = M^2(z^a) \times M^3(y^i) \times M^{1,5}(x^\mu)$, where
$M^2$ and $M^3$ are conformally flat euclidean spaces, $M^{1,5}$ is the
pseudoriemannian six-dimensional space--time and, equivalently,
the metric has the following form
\be
ds_{11}^2 = g_2^\frac12 \delta_{ab} dz^a dz^b
   + g_3^\frac13 \delta_{ij}  dy^i dy^j
   + g_2^{-\frac14} g_3^{-\frac14} g_{\mu\nu} dx^\mu dx^\nu.
   \label{MA}
\ee
Here the scalars $g_2, g_3$ and the six-dimensional metric $g_{\mu\nu}$
are functions of $x^\mu$.
The corresponding consistent truncation of $11D$
supergravity involves the following ans\"atz for the $11D$ four--form
field $\hat F_{[4]}$:
\bea
&&\hat F_{[4]\mu\nu ab} = \epsilon_{ab} F_{[2]\mu\nu},\quad
\hat F_{[4]\mu ijk} = \epsilon_{ijk} \partial_\mu \kappa, \nonumber\\
&&\hat F_{[4]\mu\nu\lambda\tau} = H_{[4]\mu\nu\lambda\tau},
   \label{FA}
\eea
where $H_{[4]}=dB_{[3]}$, $F_{[2]}=dA_{1]}$ and $\kappa$ are the
{\em six--dimensional} four--form, two--forms and pseudoscalar respectively.

It is straightforward to show that the equations
of motion for the six--dimensional variables can be derived from the
following action
\bea
S_6 &=& \int \, \sqrt{-g_6} \; d^6x \Bigl\{ R_6
   - \frac12 e^{2\phi} (\partial \kappa)^2
   - \frac12 (\partial \phi)^2
   - \frac13 (\partial \psi)^2 \nonumber \\
  &-& \frac14 e^{-\phi} \left(  e^\psi F_{[2]}^2
   + \frac1{12} e^{-\psi} H_{[4]}^2 \right) \Bigr\}
   + \int \kappa F_{[2]} \wedge H_{[4]},
     \label{EA}
\eea
where
\be
\phi = -\frac12 \ln g_3, \quad
\psi = -\frac34 \ln g_2 - \frac14 \ln g_3. \label{Phi}
\ee

Now let us start with the $8D$ Einstein theory of gravity
\be
S_8 = \int d^8x \sqrt{- g_8} R_8,
\ee
on space--times with two spacelike Killing symmetries.
In this case the $8D$ metric can be presented as standard Kaluza--Klein
pattern
\be \label{d6}
ds_8^2 = h_{ab} (d\zeta^a+\cA_{\mu}^a dx^\mu)(d\zeta^b+\cA_{\nu}^b dx^\nu)
    + (\det h)^{-\frac14} g_{\mu\nu} dx^\mu dx^\nu,
\ee
where $h_{ab}$, $2 \times 2$ matrix, and $\cA_{\mu}^a$ are two Kaluza-Klein
(KK) vectors depending on $x^\mu$ only.
Under the KK reduction to six dimensions one gets two KK two--forms,
and three scalar moduli parameterizing the metric $h_{ab}$
\bea
h_{ab} &=& e^{\frac23\psi} \left( \begin{array}{cc}
                  e^{-\phi}+\kappa^2 e^\phi & \kappa e^\phi \\
                  \kappa e^\phi & e^\phi
                  \end{array} \right), \nonumber \\
F_{[2]} &=&d \cA^1 , \quad H_{[2]} = d \cA^2, \label{DT}
\eea
forming the six-dimensional effective action
\bea
S_6 &=& \int \, d^6x \sqrt{-g_6} \Bigg\{ R_6
  - \frac12 e^{2\phi} (\partial \kappa)^2
  - \frac12 (\partial \phi)^2 - \frac13 (\partial \psi)^2 \nonumber \\
 &-& \frac14 e^{\psi} \left[ e^{-\phi} F_{[2]}^2
  + e^\phi (H_{[2]} + \kappa F_{[2]})^2 \right] \Bigg\}. \label{EB}
\eea

Note that the four--form $H_{[4]}$ from
the $11D$ supergravity is generated from the KK two--form
by inverse dualization
\be\label{DDD}
H_{[4]\alpha_1\cdots\alpha_4} = \frac12 \sqrt{-g} e^{\psi+\phi}
  \epsilon_{\alpha_1\cdots\alpha_4\mu\nu}
  \left( H_{[2]}^{\mu\nu} + \kappa F_{[2]}^{\mu\nu} \right),
\ee
after that the effective actions (\ref{EA}) and (\ref{EB}) are identical.
Therefore, any $8D$ vacuum solution with at least two spacelike Killing
vectors can be embedded into $11D$ supergravity by the following procedure:
\begin{enumerate}
\item
Writing an $8D$ Ricci flat space with two commuting Killing vectors
to the form of (\ref{d6}) and making the identifications of variables
by (\ref{DT}).

\item
Obtaining the four--form $H_{[4]}$ via the dualization (\ref{DDD}).

\item
Recovering the $11D$ metric as (\ref{MA}) through inverse variable
identification (\ref{Phi}).

\item
Deriving the $11D$ four--form field $\hat F_{[4]}$ by (\ref{FA}).
\end{enumerate}

Possible permutation of two
Killing vectors in (\ref{d6}) will lead to different $11D$ solutions.
Moreover, if the actual number of commuting spacelike isometries of the
$8D$ solution is greater than two, one can generate several different
$11D$ solutions via different choice of the ordered pair of
$\zeta^a$--directions in (\ref{d6}).

\section{$M2$-- and $M5$--Brane Solutions}
The simplest examples of application of the new embedding procedure are
the non--rotating black $M2$--brane and $M5$--brane.
Consider the $8D$ pp-wave metric
\bea
ds_8^2 &=& H \left( d\zeta_1+A_t dt \right)^2 + d\zeta_2^2
   - H^{-1} f dt^2 \nonumber\\
  &+& dx_1^2 + \cdots + dx_k^2 + f^{-1} dr^2 + r^2 d\Omega_{4-k}, \label{8d}
\eea
where $k$ in an integer, $(0 \le k \le 2)$ and
\begin{equation}
H = 1 + \frac{2\delta}{r^{3-k}}, \quad
f = 1 - \frac{2m}{r^{3-k}}, \quad
A_t = \frac{2\sqrt{\delta(m+\delta)}}{r^{3-k}H}.
\end{equation}
Two commuting Killing vector fields may be chosen either as
$(\partial_{\zeta_1}, \partial_{\zeta_2})$ or as
$(\partial_{\zeta_2}, \partial_{\zeta_1})$.
In the first case, performing
the steps described above, we arrive at the $M2$-brane solution
\begin{eqnarray}
ds_{M2}^2 &=&
  H^{-\frac23}\left( - f dt^2 + dz_1^2 + dz_2^2 \right) \nonumber\\
  &+& H^{\frac13} \Bigl( dy_1^2 + dy_2^2 + dy_3^2
   + dx_1^2 + \cdots + dx_k^2
   + f^{-1} dr^2 + r^2 d\Omega_{4-k} \Bigr), \label{BM2}
\end{eqnarray}
and $\hat A_{tz_1z_2} = A_t$.
In the second case one gets the $M5$-brane
\begin{eqnarray}
ds_{M5}^2 &=&
  H^{-\frac13} \left( -f dt^2 + dz_1^2 + dz_2^2 + dy_1^2
   + dy_2^2 + dy_3^2 \right) \nonumber\\
  &+& H^{\frac23}\left( dx_1^2 + \cdots + dx_k^2
   + f^{-1} dr^2 + r^2 d\Omega_{4-k} \right), \nonumber\\
\hat A_{x_1x_2\phi} &=& -2\sqrt{\delta(m+\delta)} \cos\theta,
  \hskip2.4cm \mbox{for } k = 2, \nonumber \\
\hat A_{x_1\phi_1\phi_2} &=& -2\sqrt{\delta(m+\delta)} \cos^2\theta,
  \hskip2.3cm \mbox{for } k = 1, \nonumber \\
\hat A_{\psi\phi_1\phi_2} &=& -2\sqrt{\delta(m+\delta)} \cos^3\theta\sin\psi,
  \hskip1.5cm \mbox{for } k = 0. \nonumber
  \label{BM5}
\end{eqnarray}
In a similar way one can obtain rotating branes, solutions endowed
with waves and KK monopoles as well as some their intersections.

The above construction illustrates that our mapping from $8D$ to $11D$
is one to many, {\em i.e.} an unique $8D$ vacuum solution can have
several $11D$ supergravity counterparts depending on the choice and
the order of two Killing vectors in eight dimensions.
This provides a simple way to proliferate $11D$ solutions: taking
one known solution one can find its $8D$ vacuum partner and then go back
to eleven dimensions with a different choice of the Killing vectors involved.
For example, two well-known five--dimensional KK
solutions, the Brinkmann wave and the KK-monopole, can be combined in a
six--dimensional vacuum solution which has six $11D$ counterparts including
all pairwise intersections of four basic solutions: $M2$--brane,
$M5$--brane, wave and monopole except the $M2 \bot M5$--brane.
One can also find the NUT--generalizations of the rotating $M$-branes in a
way similar to that used in the five-dimensional theory \cite{ChGaMaSh99}.

\section{$M2\cup M5$--Brane Solution}
A new interesting solution representing a non-standard intersection of the
$M2$ and $M5$ branes at a point (not over a string as demanded by the
usual intersection rules)
can be obtained starting with the $5D$ Kaluza--Klein dyon.
The most general dyonic black hole solution of KK
theory was found by Gibbons and Wiltshire \cite{GiWi86} and its rotating
version by Cl\'ement \cite{Cl86a}, Rasheed \cite{Ra95} and Larsen \cite{La99}.
In the static case this is a three--parameter family (with mass, electric and
magnetic charges $m, q, p$):
\bea
ds_8^2 &=& \frac{B}{A} \left( d\zeta_1 + {\cal A}_\mu dx^\mu \right)^2
   + d\zeta_2^2 + dx_1^2 + dx_2^2 \nonumber \\
  &+& \sqrt{\frac{A}{B}} \left\{ \frac{-\Delta}{\sqrt{AB}} dt^2
   + \sqrt{AB} \left( \frac{dr^2}{\Delta} + d\Omega_2 \right) \right\},
\eea
where
\bea
A &=& \left( r - \frac{d}{\sqrt{3}} \right)^2
   - \frac{2 d p^2}{d - \sqrt{3}m}, \nonumber \\
B &=& \left( r + \frac{d}{\sqrt{3}} \right)^2
   - \frac{2 d q^2}{d + \sqrt{3}m}, \nonumber \\
\Delta &=& r^2 - 2 m r + p^2 + q^2 - d^2,
\eea
and the KK vector field is given by
\be
{\cal A}_t = \frac{C}{B}, \qquad
{\cal A}_\phi = 2p\cos\theta,
\ee
with
\be
C = 2 q \left(r - \frac{d}{\sqrt{3}} \right).
\ee
The scalar charge $d$ satisfies the cubic equation
\be
\frac{q^2}{d+\sqrt{3}m} + \frac{p^2}{d-\sqrt{3}m} = \frac23 d
\ee

Using the embedding procedure, one obtains the $11D$ solution
\bea
ds_{11}^2 &=& \left( \frac{B}{A} \right)^{-\frac16} \Bigl\{
    \frac{-\Delta}{\sqrt{AB}} dt^2
  + \sqrt{AB} \left( \frac{dr^2}{\Delta} + d\Omega_2 \right) \Bigr\}
    \nonumber \\
 &+& \left( \frac{B}{A} \right)^{-\frac23}
     \left( dz_1^2 + dz_2^2 \right)
  + \left( \frac{B}{A} \right)^\frac13 \left( dy_1^2 + dy_2^2
  + dy_3^2 + dx_1^2 + dx_2^2 \right), \nonumber \\
\hat A_{tz_1z_2} &=& {\cal A}_t, \qquad
\hat A_{\phi z_1z_2} = {\cal A}_\phi. \label{M2uM5}
\eea

This solution for $p=0$  reduces to the usual $M2$--brane while for
$q=0$ --- to the $M5$--brane
(localized only on a part of the full transverse space). However, for both
$q,\,p$ non-zero, this is a new dyonic brane: contrary to the
$M2 \subset M5$--brane \cite{IzLaPaTo96,GrLaPaTo96,Co97},
for which the electric brane lies
totally within the magnetic one, and contrary to the intersecting
$M2 \bot M5$--brane (with a common string), now the common part
of the world-volumes is zero-dimensional (we suggest the name
$M2 \cup M5$).
\[ \begin{array}{rccccccc}
{M2}: & z_1 & z_2 &     &     &     &     &     \\
{M5}: &     &     & y_1 & y_2 & y_3 & x_1 & x_2 \\
\end{array} \]
The electric/magnetic charge densities are $Q=8\pi q$
and  $P=8\pi p$ respectively.

One can also obtain the rotating version of $M2\cup M5$--brane
(with one rotational parameter) through the following transformation:
\bea
dt &\to& dt + \omega d\phi, \qquad
  \Delta \to \Delta + a^2,  \nonumber \\
A &\to& A + a^2\cos^2\theta
  + \frac{2 j p q \cos\theta}{(m+\frac{d}{\sqrt{3}})^2 - q^2},
  \nonumber \\
B &\to& B + a^2\cos^2\theta
  - \frac{2 j p q \cos\theta}{(m-\frac{d}{\sqrt{3}})^2 - p^2},
  \nonumber \\
C &\to& C - \frac{2 j p (m+\frac{d}{\sqrt{3}}) \cos\theta}
                 {(m-\frac{d}{\sqrt{3}})^2 - p^2},
\eea
where
\be
\omega = \frac{2j\sin^2\theta}{\Delta-a^2\sin^2\theta} \Bigl[ r-m
 + \frac{(m+\frac{d}{\sqrt{3}})(m^2+d^2-p^2-q^2)}
           {(m+\frac{d}{\sqrt{3}})^2-q^2} \Bigr].
\ee
In terms of new variables one component of the KK vector potential,
$\cA_t$, remains the same  while another becomes
\bea
\cA_\phi &=& \frac{C}{B}\omega
  + \frac{2p\Delta\cos\theta}{\Delta-a^2\sin^2\theta} \nonumber \\
 &-& \frac{2jq\sin^2\theta \left[ r(m-\frac{d}{\sqrt{3}})+\frac{md}{\sqrt{3}}
           - p^2 - q^2 + d^2 \right]}
    {(\Delta-a^2\sin^2\theta)\left((m+\frac{d}{\sqrt{3}})^2-q^2 \right)}.
\eea
The angular momentum of the system, $j$, is given by
\be
j^2 = a^2 \frac{\left[ (m+\frac{d}{\sqrt{3}})^2-q^2 \right]
      \left[ (m-\frac{d}{\sqrt{3}})^2-p^2 \right]}
                {m^2 + d^2 - p^2 - q^2}.
\ee

\section{Conclusion}
We have presented a new generating technique to obtain solutions to
$11D$ supergravity with non-zero four-form field starting with the
solutions of vacuum 8D gravity with two commuting Killing spacelike Killing
symmetries. This method can also be combined with other tools
to get physically interesting solutions. In particular, a direct
application of our procedure does not allow to obtain $11D$ branes
{\em localized} on all coordinates of transverse space, since
our basic ansatz (\ref{MA}) is not general enough. However one still can use
it to explore possible geometric $11D$ structures and then to try to localize
solutions in a straightforward way.
Thus we have given a purely vacuum interpretation to a certain class
of the $11D$ supergravity $M$-branes via new non-local duality between $11D$
supergravity and $8D$ vacuum Einstein gravity. It is worth noting that
the very natural correspondence between the Killing spinor equations
in both theories may have a deeper meaning in the context of $8D$
supergravities \cite{ChGaSh99a}.
In this connection it has to be emphasized that the type
of duality described here is by no means related to the direct dimensional
reduction from $11D$ to $8D$ theory. It is also different from the duality
between the $8D$ self-dual gauge theories and $10D/11D$ supergravities
discussed in \cite{IzLaPaTo96,GrLaPaTo96} where the four-form has
a non-geometric origin.

\section*{Acknowledgments}
The work of CMC was supported in part by the National Science Council
under grant NSC 89-2112-M-008-016.

\end{document}